\documentclass[twocolumn,showpacs,preprintnumbers,prd,floatfix]{revtex4} 
\usepackage{epsfig}
\usepackage{array}
\usepackage{graphicx}
\usepackage{dcolumn}
\usepackage{bm}
\usepackage{amssymb,amsmath}
\usepackage{natbib}
\usepackage{url}

\def\beq{\begin{equation}}
\def\eeq{\end{equation}}
\def\bea{\begin{eqnarray}}
\def\eea{\end{eqnarray}}

\begin{document}
\title{Finite Mirror Effects in Advanced Interferometric Gravitational Wave Detectors}
\author{Andrew P. Lundgren}
\email{apl27@astro.cornell.edu}
\affiliation{Department of Physics, Cornell University, Ithaca, New York 14853}
\affiliation{Center for Radiophysics and Space Research, Cornell University, Ithaca, NY 14853, USA}
\author{Ruxandra Bondarescu}
\email{ruxandra@astro.cornell.edu}
\affiliation{Department of Physics, Cornell University, Ithaca, New York 14853}
\affiliation{Center for Radiophysics and Space Research, Cornell University, Ithaca, NY 14853, USA}
\author{David Tsang}
\email{dtsang@astro.cornell.edu}
\affiliation{Department of Physics, Cornell University, Ithaca, New York 14853}
\affiliation{Center for Radiophysics and Space Research, Cornell University, Ithaca, NY 14853, USA}
\author{Mihai Bondarescu}
\email{mihai@aei.mpg.de}
\affiliation{Max Planck Institut f\"ur Gravitationsphysik, Albert Einstein Institut, 14476 Golm, Germany}
\affiliation{California Institute of Technology, Pasadena, CA 91125, USA}
\date{\today}
\begin{abstract}
Thermal noise is expected to be the dominant source of noise in the most sensitive frequency band of second generation ground based gravitational wave detectors. Reshaping the beam to a flatter wider profile which probes more of the mirror surface reduces this noise.  The ``Mesa'' beam shape has been proposed for this purpose and was subsequently generalized to a family of hyperboloidal beams with two parameters: twist angle $\alpha$ and beam width $D$. Varying $\alpha$ allows a continuous transition from the nearly-flat ($\alpha = 0$) to the nearly-concentric ($\alpha = \pi$) Mesa beam configurations. We analytically prove that in the limit $D \to \infty$ hyperboloidal beams become Gaussians. The ideal beam choice for reducing thermal noise is the widest possible beam that satisfies the Advanced LIGO diffraction loss design constraint of 1 part per million (ppm) per bounce for a mirror radius of 17 cm. In the past the diffraction loss has often been calculated using the clipping approximation that, in general, underestimates the diffraction loss. We develop a code using pseudo-spectral methods to compute the diffraction loss directly from the propagator. We find that the diffraction loss is not a strictly monotonic function of beam width, but has local minima that occur due to finite mirror effects and leads to natural choices of $D$. For an $\alpha = \pi$ Mesa beam a local minimum occurs at $D = 10.67$ cm and leads to a diffraction loss of $1.4$ ppm.  We then compute the thermal noise for the entire hyperboloidal family. We find that if one requires a diffraction loss of strictly 1 ppm, the $\alpha = 0.91 \pi$ hyperboloidal beam is optimal, leading to the coating thermal noise (the dominant source of noise for fused-silica mirrors) being lower by about 10\% than for a Mesa beam while other types of thermal noise decrease as well. We then develop an iterative process that reconstructs the mirror to specifically account for finite mirror effects. This allows us to increase the $D$ parameter to $11.35$ cm for a nearly-concentric Mesa beam and lower the coating noise by about 30\% compared to the original Mesa configuration.  
\end{abstract}

\maketitle

\section{Introduction}
The initial baseline design for the Advanced LIGO gravitational wave detectors \cite{LIGO, StandardLIGOreference} 
employs Gaussian beams in the arm cavities.  The leading noise source in the most sensitive frequency band of the instruments ($\sim 30-300$ Hz) is the thermal noise in the substrate and reflective coating of the mirror test masses.  Lowering thermal noise is therefore of paramount importance for achieving a higher event rate in LIGO.  There are a number of other detectors that are being built or upgraded to similar specifications.  While we will choose to study Advanced LIGO for definiteness, our general conclusions should be more widely applicable to any interferometeric detector that needs to limit thermal noise.  Some of the important parameters that we use are summarized in Table \ref{AdvLIGOParams}.

LIGO is a Fabry-Perot interferometer with four mirrored test masses. The resonant beams in the cavity measure the position of the test masses, averaging over the mirrored surface, with the average weighted by the power distribution of the beam. Thus, the highly illuminated central area is weighted more than the mirror boundary that is left nearly dark. One way of decreasing the thermal noise is to flatten the beam so that a larger fraction of the mirror is in use. Motivated by this intuitive reasoning, O'Shaughnessy et. al. \cite{D'Ambrosio:2004wz, O'Shaughnessy:2004qh} proposed the flat topped Mesa beams, which were subsequently explored in detail by them and others \cite{Agresti:2005ad, Savov:2004nv,MihaiThesis,DAmbrosio}.  These beams would lower the thermal noise by a factor of approximately 2.5 compared to the baseline design. The original Mesa beam supported by nearly-flat Mexican Hat mirrors was found to be susceptible to a tilt instability \cite{Sidles}.  This triggered the proposal of a Mesa beam supported by nearly-concentric mirrors \cite{BT}.  In the same paper, a family of hyperboloidal beams that include all Mesa and Gaussian beams previously considered was introduced.  Mesa is currently the leading alternative beam design for Advanced LIGO, and is being studied experimentally \cite{Experiment1, TaralloThesis}.

In this paper we first discuss the formulation of hyperboloidal beams.  The ``nearly-flat'' Mesa is created by superposing minimal Gaussians with generators uniformly distributed inside a cylinder, and the ``nearly-concentric'' by generators falling inside a cone, and passing through the center of the cavity.  These two choices have the same intensity distribution on the mirrors, but the second has a much smaller susceptibility to tilt instability.  The hyperboloidal beams smoothly interpolate between these two cases by twisting the generators of the minimal Gaussians by an angle $\alpha$.  After discussing some geometric properties of the beams, we present a proof that Gaussian beams are a special case of the hyperboloidal beams, confirming a conjecture in \cite{BT}.

We then compute the three types of mirror thermal noise for a variety of hyperboloidal beam shapes, using a set of simple scaling laws developed in \cite{Lovelace,OShaughnessy} that simplify previous work \cite{Levin,LiuThorne,AgrestiDeSalvoNoise}.  The first is substrate Brownian noise, occuring due to mechanical dissipation in the material; this is the least significant source of thermal noise.  The substrate thermoelastic noise is caused by random thermal expansion.  The coating also has both Brownian and thermoelastic noise, but these follow the same scaling laws so we consider them as a single source.  The coating noise is the most severe of the three types for the fused-silica substrates currently planned for Advanced LIGO.  The substrate thermoelastic noise would dominate in a material like sapphire which has a higher thermal expansion coefficient.  We show that the noises decrease with increasing width of the beam, as expected, and that the hyperboloidal beams have larger noise than the relevant Mesa beams.

The constraint on our ability to lower the noise comes from the need to keep the diffraction loss small.  Gravitational-wave interferometers must keep a large circulating power in the cavity, and so cannot allow significant amounts of light to escape past the edge of the mirrors.  The current design constraint used in the most recent papers \cite{PierroGaldi:2007,MihaiYanbei,MihaiThesis} is a diffraction loss of 1 part per million (ppm) per bounce for 17 cm fused-silica mirrors.  The clipping approximation Eq. \eqref{ClippingApproximation} indicates that the desired Mesa width is approximately $D = 10$ cm.  Previous work \cite{Barriga, DAmbrosio} have shown that the clipping approximation is not accurate for Gaussian beams of finite mirrors, and have used Fast Fourier transform simulations for accurate calculations.

We calculate the diffraction losses accurately from eigenvalues of the cavity propagator using an exponentially convergent code that enables us to study the structure of Mesa and other hyperboloidal beams as a function of $D$, mirror radius, and twist angle $\alpha$ in detail.  We find that the diffraction loss is not a monotonic function of $D$, but due to finite mirror effects has anomalous local minima where the loss is significantly below what is expected from the clipping approximation.  These minima are observed to become more shallow and eventually disappear as the radius of the mirror is increased.  However, for the mirror radii and beam widths relevant for Advanced LIGO the finite mirror effects are important.  We show that they can be used to increase the width of the beam, lowering the noise even further than previous work.  
 
Finally, we develop an iteration scheme to redesign the mirror, explicitly accounting for finite mirror effects.  The iterated mirror is altered to match the phasefront of the primary eigenbeam of the finite mirror cavity, reducing the diffraction loss of this mode, thus allowing even larger beam widths to satisfy the diffraction loss constraint.

The mathematical construction of the hyperboloidal beams are discussed in Sec. II, while the asymptotic limit of the hyperboloidal beams are derived in Sec. III. The thermal noise scaling laws are described in Sec. IV. The cavity propagator construction and eigenmode decomposition are presented in Sec. V, with the results, including finite mirror effects, discussed in Sec VI. We then summarize our work in Sec. VII.

\begin{table}[h]
\begin{center}
\begin{tabular}{|c|l|l|}\hline
$L$ & $3999.01$ m & Length of LIGO cavity\\
$~~~\lambda_0~~~$ & $1.064\times 10^{-6}$ m & Laser wavelength\\
$w_o$& $\sqrt{\lambda_0 L / 2\pi} = 2.6023$ cm & Minimal Gaussian width\\
$R$ & $17$ cm & Mirror radius \\
\hline
\end{tabular}
\caption{Advanced LIGO Parameters}
\label{AdvLIGOParams}
\end{center}
\end{table}

\section{Construction of the Beams}

\begin{figure}
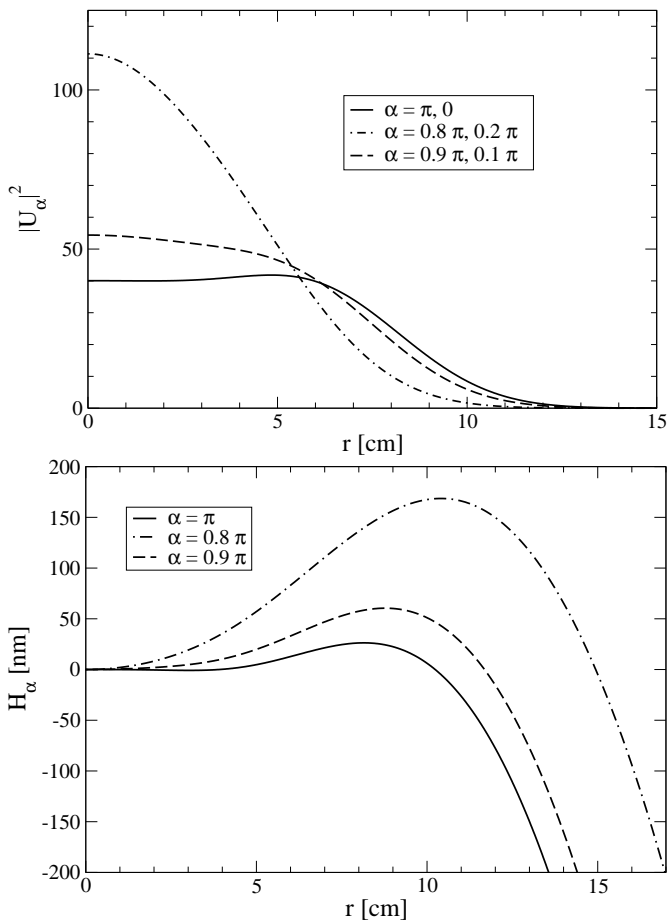

\begin{center}
\leavevmode
\epsfxsize=250pt
\epsfbox{fig1a.eps}
\epsfxsize=250pt
\epsfbox{fig1b.eps}
\caption{(a) The beam intensity profile $|U_\alpha|^2$ and (b) corrections $H_\alpha$ are shown at fixed $D = 10$ cm for different twist parameters $\alpha$. }
 \renewcommand{\arraystretch}{0.75}
 \renewcommand{\topfraction}{0.6}
  \label{beamIntensity}
\end{center}
\end{figure}

The beams we study are supported by two identical mirrors facing each other, forming a cavity of length $L$.  The mirrors are cylindrically symmetric around the optical axis, which runs along the length of the cavity and will be called the $z$ axis.  The center of the cavity, equidistant between the mirrors, is $z = 0$; the mirrors are located at $z = -z_R$ and $z = z_R$, where $z_R = L / 2$.  The transverse distance from the $z$ axis will be denoted by $r$, and the angular coordinate by $\phi$.

The cavity is fed with laser light with wavelength $\lambda_0$, and the distance between the mirrors is fine-tuned so that the cavity resonates in its fundamental mode, with a field amplitude $U(r,z)$ and intensity $|U(r,z)|^2$.  In this paper we focus on axisymmetric modes with no $\phi$ dependence in the beams.  Non-axisymmetric modes are important for studies of the tilt and parametric instabilities \cite{Savov:2004nv,PI1}, but are not discussed in detail in this work.  The narrowest possible Gaussian mode that can exist in a cavity of given length is called the minimal Gaussian, which has the intensity distribution
\beq
|U(r,z)|^2 = \frac{2}{1+(z^2/z_R^2)} \exp\left(-\frac{2 r^2}{w_0^2 (1+(z^2/z_R^2))}\right) ~,
\eeq
where $w_0 = \sqrt{\lambda_0 z_R / \pi}$.

A hyperboloidal beam is the superposition of minimal Gaussians chosen such that the symmetry axis of the individual minimal Gaussians are generators of a set of coaxial hyperboloids.  The beam family has two parameters: $\alpha$, the twist angle one would have to rotate the two basis of a set of coaxial cylinders with respect to each other to obtain the hyperboloids and $D$, the radius of a section perpendicular to the optic axis of the outermost hyperboloid at the end of the cavity. In the case $\alpha = 0$, the propagation axes are parallel and fill a cylinder of radius $D$.  This is the Mesa beam supported by nearly-flat mirrors.  For $\alpha = \pi$ the lines all cross at $z = 0$ forming two cones.  This configuration also generates a Mesa beam, but one supported by nearly-concentric mirrors.  Varying $\alpha$ smoothly deforms the beam and the mirror shape between the two configurations.  Some examples of the beam shapes are displayed in Fig.\ \ref{beamIntensity}a.

For the cavity to support the desired beams, the phase of the electric field of the beam should be constant on the mirror surface.  We will focus our attention on the mirror on the positive $z$ side of the cavity.  The wavefront can be approximated by the ``fiducial spheroid'',
\beq
z = S_\alpha(r) = \sqrt{z_0^2 - r^2 \sin^2 (\alpha/2)} ~.
\eeq
For $\alpha = 0$ this is the mirror plane $z = z_R$; for $\alpha = \pi$, the fiducial spheroid is a sphere centered on $z = 0$, and clearly the lines which generate the hyperboloid are all radii of the sphere.

There are two equivalent expressions for the field amplitude evaluated on the fiducial spheroid.  The first is the integral expression \cite{BT}
\bea
U_\alpha(r,S_\alpha) &=& \Lambda \int^{R_0}_0 \int^{2\pi}_0 d\phi_0 r_0
\exp\left[ i\frac{r r_0}{w_0^2} \sin \phi_0 \sin \alpha  \right.  \nonumber \\
&-& \left.  \frac{(r^2 + r_0^2-2r r_0 \cos \phi_0)}{2 w_0^2}(1 - i \cos \alpha) \right] ~,
\eea
where $\Lambda$ is a complex constant.

The second is the method that we use in this paper.  The beam is constructed as shown in \cite{Galdi:2006rq} by an expansion in Gausse-Laguerre eigenbeams of spherical mirrors.  They are closely related to the Gauss-Laguerre basis functions given by
\beq
\psi_m(\xi) = \sqrt{2} \exp (-\xi^2/2) L_m(\xi^2) ~,
\eeq
where $L_m$ is the $m^{th}$ Laguerre polynomial.  Then the Gauss-Laguerre eigenbeams are
\bea
\Psi_m(r,z) = \frac{w_0}{w(z)} ~ \psi_m\left[\frac{\sqrt{2}r}{w(z)}\right] ~ \exp\left[\frac{i k_0 r^2}{2R(z)}\right] 
\\ \nonumber \times \exp[i(k_0 z - (2m+1)\Phi(z))]
\eea
where
\bea
w(z) = w_0 \sqrt{1+(z/z_R)^2} ~, \quad R(z)=z+z_R^2/z , \\ \nonumber
 \quad \Phi(z) = \arctan (z / z_R )
\eea
and $k_0 = 2 \pi / \lambda_0$.  The expansion is written as
\beq
U_\alpha (r, z) = \sum_{m=0}^{\infty} A_m^{(\alpha)} \Psi_m (r, z) ~.
\eeq
The expansion coefficients that result in hyperboloidal beams are
\beq
A_m^{(\alpha)} = (- \cos \alpha)^m \frac{\sqrt{2} w_0^2}{D^2} P\left( m+1, \frac{D^2}{2 w_0^2} \right) ~.
\eeq
$P(a,x)$ is the incomplete gamma function
\bea
P(a,x) &=& \frac{1}{\Gamma(a)} \int_0^x e^{-t} t^{a-1} dt  \\ \nonumber
&=& \frac{ \int_0^x e^{-t} t^{a-1} dt}{ \int_0^\infty e^{-t} t^{a-1} dt} ~.
\eea

The mirror shape that supports a hyperboloidal beam is not exactly the fiducial spheroid.  We make a correction $h(r)$ so that the surface of the mirror is given by $z_M(r) = S_\alpha(r) - h(r)$.  The correction is chosen so that the mirror is located at a phasefront of the beam.  We find the mirror surface from $U_\alpha(r,S_\alpha)$ by
\beq
h(r) = k_0^{-1} ( U_\alpha(r,S_\alpha) - U_\alpha(0,S_\alpha) ) ~.
\eeq
The shape of $h(r)$, or the mirror itself, is generally referred to as a ``Mexican hat'', and some examples are displayed in Fig.\ \ref{beamIntensity}b.

As expected, the beams for $\alpha$ and $-\alpha$ are identical, as they correspond to hyperboloids that are simply twisted in opposite directions.  There is a duality between $\alpha$ and $\pi-\alpha$.  This was first mentioned in \cite{BT} for all $\alpha$'s, then studied in more depth for $\alpha=0$ and $\alpha = \pi$ in \cite{Savov:2004nv} and, finally, analytically understood in \cite{Agresti:2005ad}.  This duality extends to several quantities.  The beam intensity profiles are identical.  The corrections to the mirror shape are opposite, $h_\alpha(r) = - h_{\pi-\alpha}$.  There are also dualities in the eigenvalues of the propagator \cite{Savov:2004nv,Agresti:2005ad,Galdi:2006rq,BT} that we will not discuss in this paper.

\section{Asymptotic Behavior of Wide Hyperboloidal Beams}

It was conjectured by Bondarescu and Thorne \cite{BT} that the beam becomes a Gaussian in the limit $D \to \infty$.  We will prove this analytically for the intensity profile of the beam, evaluated on the plane $z = z_R$ which would be the surface of a perfectly flat mirror.  The intensity varies slowly enough with $z$ that this will also be the intensity profile on the mirror to a good approximation.  Our proof uses the expression for the beam amplitude in terms of a summation of Gauss-Laguerre functions.  The essential ingredient is the realization that the expansion coefficients take the form $A^{(\alpha)}_m = (constant)^m$ as $D \to \infty$, where the constant depends only on $\alpha$.  This allows us to analytically perform the summation to obtain the beam profile.

In the limit $x \rightarrow \infty$, the incomplete gamma function $P(a,x) = 1$, giving $A_m =  \sqrt{2} w_0^2 (- \cos \alpha)^m / D^2$.  The approximation $z = z_R$ yields
\beq
w(z_R) = \sqrt{2} w_0 ~, \quad R(z_R) = 2 z_R ~, \quad \Phi(z_R) = \pi / 4 ~,
\eeq
and the Gauss-Laguerre propagators become
\beq
\Psi_m(\bar r,z_R) = \frac{1}{\sqrt{2}} \psi_m(\bar r) e^{- i \frac{\pi}{2} m} e^{i \phi(\bar r)} ~,
\eeq
where $\bar r = r /w_0$ and the r-dependent part of the phase has been absorbed into $\phi(\bar r)$.  Since in the end we will only be interested in the intensity profile, the exact form of $\phi(\bar r)$ is unimportant.

The expansion for $U_\alpha$ then becomes
\bea
U_\alpha(\bar r,z_R) &=& \sum_{m=0}^\infty (- \cos \alpha)^m \frac{w_0^2}{D^2} \left( e^{-\frac{i \pi}{2}} \right)^m e^{-\frac{\bar r^2}{2}} L_m(\bar r^2) e^{i \phi(\bar r)} \nonumber \\
&=& \left(\frac{w_0}{D}\right)^2 e^{- \frac{\bar r^2}{2}} e^{i \phi(\bar r)} \sum_{m=0}^{\infty} (i \cos \alpha)^m L_m(\bar r^2)
\label{AmplitudeAnalyticProof}~.
\eea
We now use the generating function for the Laguerre polynomials \cite{AbramowitzStegun}
\beq
\sum_{m=0}^{\infty} L_m (x) t^m = \frac{1}{1 - t} \exp\left(- \frac{t}{1 - t} x \right)
\eeq
to evaluate the sum in \eqref{AmplitudeAnalyticProof}, with $t = i \cos \alpha$.  The final result is an intensity profile
\beq
| U_\alpha(r) |^2 = \frac{1}{\pi w_0^2} \exp \left[ - \frac{r^2}{\sigma^2} \right] ~, \quad \sigma = \frac{w_0 \sqrt{1 + \cos^2 \alpha}}{\sin \alpha} ~.
\label{analyticGW}
\eeq

The minimal Gaussian $\alpha = \pi / 2$ is seen to have $\sigma = w_0$.  The width is symmetric under $\alpha \rightarrow \pi - \alpha$, as expected from the duality relation \cite{BT, Agresti:2005ad}, and goes to infinity at $\alpha = 0$ or $\pi$.  This includes every Gaussian beam capable of resonating in a cavity of the given length.

\section{Thermal Noises}
There are a number of noise sources limiting the sensitivity of ground-based gravitational-wave interferometers.  Seismic noise causes an effective cutoff in the lowest frequencies that can be measured.  Fundamental problems such as shot noise and radiation pressure noise, as well as technical issues, are important limitations on the sensitivity throughout the frequency band.  However, the major contribution in the most sensitive frequencies of LIGO is the thermal noise in the mirrors.  Reducing the thermal noise is the goal of this paper.

The mirror consists of a substrate with a coating, and we must consider noises due to fluctuations of both.  The substrate and coating have both thermoelastic and Brownian contributions to the noise.  Thermoelastic noise is caused by expansions in the material caused by random heat flow.  Brownian noise is due to the coupling of normal modes of vibration by imperfections in the material.  As a practical matter, both types of noises in the coating have the same scaling law so they do not have to be considered seperately.  In fact, for the fused silica mirrors now under consideration for Advanced LIGO, the coating noises are the dominant contribution.  However, use of a material like sapphire, with a higher coefficient of thermal expansion, would cause the substrate thermoelastic noise to dominate.  We will calculate all three types of noise in this paper.

A set of simple scaling laws were derived by Lovelace \cite{Lovelace} in parallel with O'Shaughnessy \cite{OShaughnessy} that are applicable to beams of arbitrary shape.  The noise is proportional to an integral depending on the intensity and an overall constant which is independent of the shape of the beam.  The noises are given by
\beq
S_n = C_n \int_0^\infty \tilde I(k)^2 k^n dk ~, \label{GLNoise1}
\eeq
\beq
\tilde I(k) = \int_0^R J_0(k r) |U(r)|^2 r dr ~,
\eeq
where $n$ specifies the type of noise under consideration, and $\tilde I(k)$ is the 2D axisymmetric Fourier transform of the beam intensity with $k$ the radial wavenumber.  The substrate Brownian noise has $n = 0$, coating Brownian and coating thermoelastic noises have $n = 1$, and substrate thermoelastic noise has $n = 2$.  We are interested in comparing noises given by different beam shapes, so the overall constants $C_n$ are not important.  The resulting amplitude sensitivity is the square root of the noise, and has units of $\rm{meters} / \sqrt{Hz}$.

These scaling laws were derived for half-infinite mirrors, meaning that effects of the finite radius and thickness of the mirror are ignored.  For the specific mirrors under consideration for Advanced LIGO, the width and thickness of the mirror are large enough compared to the beam width \cite{Lovelace} that this should be a good approximation.  The results of \cite{AgrestiDeSalvoNoise} suggest that we can expect corrections of not much more than ten percent to the half-infinite scaling law expressions that we are using.  

\section{Eigenvalues of the Propagator}

The idealized picture of a locked cavity is that the mirrors are perfectly aligned and a precise distance from each other.  The beam should leave one mirror, reflect off the other, and when it returns it should be the same shape and exactly in phase.  The beam will have lost some intensity due to diffraction and the finite extent of the mirror.  In order to build up a very intense beam with a relatively weak laser, the beam must reflect very many times.  The loss per half-trip (from one mirror to the other) must be below approximately 60 ppm.  The majority of the loss will be due to absorption in the mirror and other factors, not diffraction.  The commonly agreed upon budget for losses due to diffraction is 1 ppm.

In previous work, the clipping approximation is often used to estimate the diffraction loss by calculating the fraction of the intensity of the beam which falls outside the mirror.  It is given by
\beq
\mathcal{DL} \approx 2 \pi \int_R^\infty |U_\alpha(r)|^2 r dr ~. \label{ClippingApproximation}
\eeq
Our numerical code is accurate and fast enough to compute the diffraction loss directly from the eigenvalues of the propagator, which is more accurate.  

The propagator also allows us to estimate the difficulty of locking the interferometer.  The finesse of Advanced LIGO is about 1200 \cite{AdvLIGOFinesse}.  This sets the width of the resonance for the cavity \cite{Saleh} to be about $2 \pi / 1200 \approx 0.005$ radians.  If any other modes with a small diffraction loss have an argument within this distance of the desired mode, there will be severe problems with locking the cavity.  For axisymmetric modes, we find that this is not the case and the arguments of the eigenvalues are well-separated enough that locking with Mesa or any of the other beams that we study should be no more difficult than locking the currently proposed Mesa.

\subsection{Integral Form of the Propagator}
The propagator from a single point $r,\phi$ on one mirror to a point $r',\phi'$ on the other (see for instance \cite{Agresti:2005ad}) is
\beq
\mathcal{K}(r,\phi,r',\phi') = \frac{i k_0}{4 \pi \rho} (1 + \cos \theta) e^{- i k_0 \rho} ~,
\eeq
where $\rho$ is the path length between the two points, $\theta$ the angle between the cavity axis and the path, and $r$ and $\phi$ are the standard cylindrical radial coordinate and azimuthal angle.  The cavity is very long compared to the radius of the mirrors so we can immediately make the paraxial approximation $\theta = 0$.  The path length can be approximated as
\bea
\rho &=& ( ( S_\alpha(r) + S_\alpha(r') - h(r) - h(r') )^2 \\ \nonumber
  &+& (r + r' - 2 r r' \cos (\phi-\phi'))^2 )^{1/2} \\ \nonumber
&\approx& L + \cos \alpha \left( \frac{r^2}{2 L} + \frac{r'^2}{2 L} \right) - h(r) - h(r') \\ \nonumber
&-& \frac{r r'}{L} \cos (\phi - \phi')
\eea
since $L \gg r$ and $L \gg h(r)$.  This expression is the path length used to find the phase of the propagator.  In the prefactor of the propagator we only need the much simpler $\rho \approx L$ in order to compute the amplitude to an accuracy of $R_{mirror}^2 / L^2 \sim 10^{-9}$. 

We assume that the cavity is axisymmetric, which means that the eigenmodes of the propagator can also be written as eigenstates of rotation.  Thus we write the complex beam amplitude as $V^{(i)}_m(r,\phi) = V^{(i)}_m(r)~e^{-i m \phi}$.  To determine the modes of the cavity, we have an eigenvalue problem given by the integral
\beq
\gamma^{(i)}_m V^{(i)}_m(r,\phi) = \int_0^R r' dr' d\phi' ~ \mathcal{K}(r,\phi,r',\phi') V^{(i)}_m(r',\phi') ~,
\eeq
where $\gamma^{(i)_m}$ is the associated eigenvalue, and $R$ is the mirror radius.  Integrating over $\phi'$ produces
\beq
\gamma^{(i)}_m V^{(i)}_m(r) =  \int_0^R r' dr' \mathcal{K}^r_m(r,r') V^{(i)}_m(r') ~, \label{AxiEigenvector}
\eeq
\bea
\mathcal{K}^r_m(r,r') &\equiv& \frac{i^{m+1} k_0}{L} J_m\left(\frac{k_0 r r'}{L}\right)
\exp \left[ \frac{}{} i k_0 \left( - L + h(r) \right. \right. \nonumber\\
& & \left. \left. + h(r') - \cos \alpha \left( \frac{r^2}{2 L} + \frac{r'^2}{2 L} \right) \right) \right] \nonumber  ~, \\
& &
\eea
where for convenience we have defined a ``radial kernel'' $\mathcal{K}^r_m$.  $J_m(x)$ is the $m^{\rm th}$ order Bessel function of the first kind.  In what follows, we specialize to axisymmetric modes ($m = 0$), as we are not focusing on tilt or parametric instabilities which involve modes with $m > 0$.

\subsection{Discrete Form of the Propagator}
The above integral must be converted to a discrete sum to be suitable for numerical computation.  We choose to do this using a Chebyshev quadrature \cite{AbramowitzStegun, Boyd}, which is appropriate for the finite range of $r'$.  We use $N$ collocation points $r_k$ with associated integration weights $w_k$.  The integral is converted to
\beq
\gamma^{(i)} V^{(i)}_0(r_j) = \Sigma_{k=1}^{N} \mathcal{K}^r_0(r_j,r_k) r_k w_k V^{(i)}_0(r_k)~.
\eeq
This is a matrix eigenvalue problem which is easy to solve numerically.  We will order the eigenstates by the number of radial nodes, the number of times that $V^{(i)}_0(r)$ goes to zero.  The fundamental mode ($i = 0$) has zero nodes and as $R \to \infty$ it limits to the hyperboloidal beam that the cavity is designed to support.  We will show that finite mirror effects cause the actual eigenstate to be slightly different.

We performed a convergence test where we varied the number of collocation points, $N$, and observed exponential convergence. The diffraction loss for the lowest eigenvalue for the $\alpha = \pi$ Mesa beam with $D = 10.67$ cm, $R = 17$ cm changes in the 5th digit when varying $N$ between $N=250$ and $N=500$.  We are using $N = 1000$ or $N = 500$ for all of the calculations in this paper.  On a typical single processor laptop computer, our code takes $\sim 10$ seconds for $N = 500$ and $\sim 1$ minute for $N = 1000$, to calculate the eigenmodes of any given cavity.

\subsection{Meaning of Eigenvalues}
The eigenvectors of the propagator are the field amplitudes of the cavity's resonant modes.  Fine-tuning the length $L$ of the cavity to $L + \delta L$ changes the eigenvalue to $\exp(- i k_0 \delta L) \gamma$.  This tuning is used to select the desired mode; when the argument of the eigenvalue is zero it will resonate in the cavity.  The magnitude of the eigenvalue is the fraction by which the amplitude changes during a half-trip.  We must have $|\gamma| < 1$ for finite mirrors, because some light will always be lost to diffraction.  We define the diffraction loss in parts per million (ppm) over one half trip through the cavity as
\beq
\mathcal{DL} = 10^6 (1 - |\gamma|^2) ~. \label{DiffLossLambda}
\eeq
Advanced LIGO requires a diffraction loss per half trip of about one ppm.  In the next section, we will discuss the results of our analysis, where we study the noise characteristics of the hyperboloidal beams given the constraint on the diffraction loss.

\section{Results}

\begin{figure}
\begin{center}
\leavevmode
\epsfxsize=235pt
\epsfbox{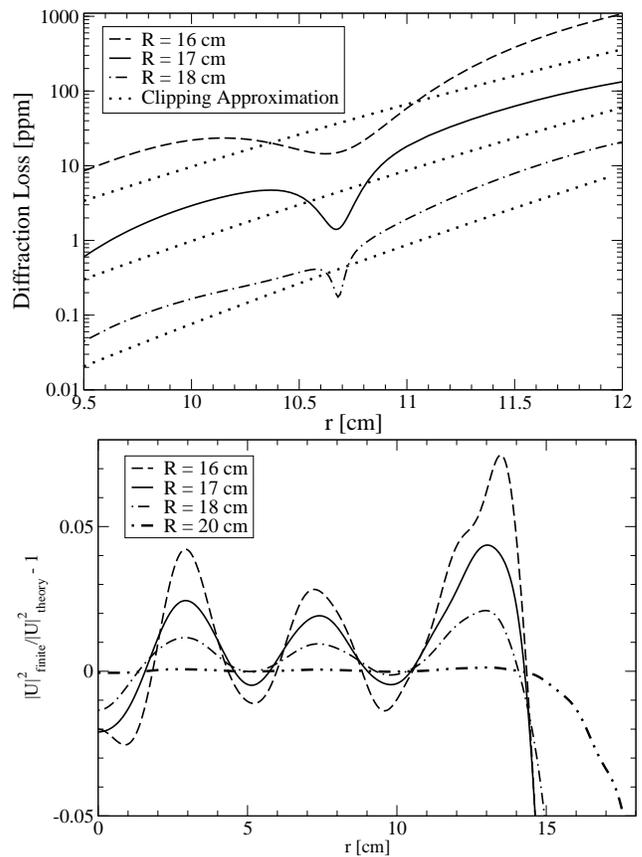}
\newline
\epsfxsize=235pt
\epsfbox{fig2b.eps}
\caption{ (a) The diffraction loss for an ($\alpha = \pi$) Mesa configuration ($\alpha = \pi$) is shown as a function of $D$ on a logarithmic plot for several different mirror radii: $R = 16$ cm, $R = 17$ cm and $R= 18$ cm. The diffraction loss computed numerically using Eq. \eqref{DiffLossLambda} (solid, dashed and dotted-dashed lines) exhibits local minima due to finite mirror effects. It can be seen that the minima get narrower as $R$ increases and that they go below the values estimated using the clipping approximation (dotted lines). (b) The fractional difference $|U|^2_{\rm finite} / |U|^2_{\rm theory} - 1$ between the theoretical infinite-mirror beam intensity profile and the actual profile given by the first eigenvector of the propagator in Eq. \eqref{AxiEigenvector} is plotted for $D = 10.67$ cm and $R = 16, 17,18$ and $20$ cm.  The deviation decreases with $R$ as expected. }
 \renewcommand{\arraystretch}{0.75}
 \renewcommand{\topfraction}{0.6}
  \label{DLossvsR}
\end{center} 
\end{figure}

\begin{figure}
\begin{center}
\leavevmode
\epsfxsize=235pt
\epsfbox{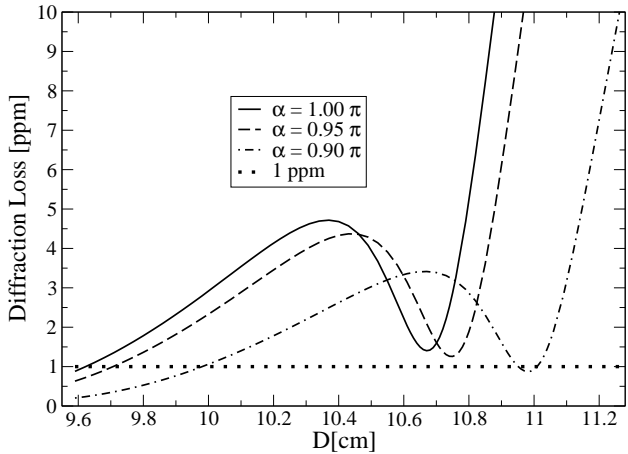}
\caption{Diffraction loss as a function of $D$ is displayed for $\alpha=\pi$, $\alpha = 0.95 \pi$, $\alpha = 0.90 \pi$. It can be seen that as $\alpha$ decreases the minimum diffraction loss is lower and occurs for larger $D$.}
 \renewcommand{\arraystretch}{0.75}
 \renewcommand{\topfraction}{0.6}
  \label{DLossvsD}
\end{center} 
\end{figure}

\begin{figure}
\begin{center}
\leavevmode
\epsfxsize=235pt
\epsfbox{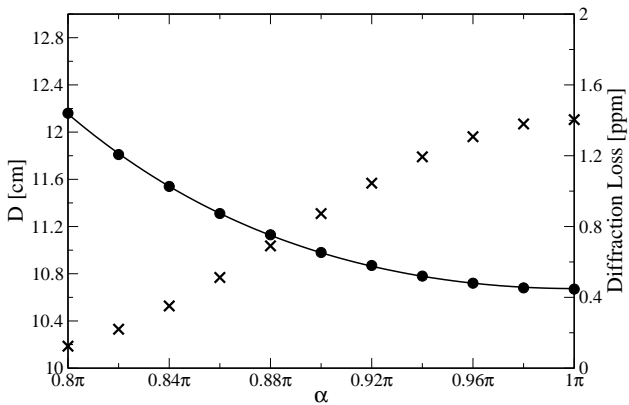}
\caption{The minimum diffraction loss in ppm (represented by crosses) and the corresponding $D$ (dots) are shown as a function of $\alpha$. The solid line represents the best exponential fit of the form $a+ \exp(b + c \sin^2\alpha)$ with $a = 0.094$, $b=-4.34$ and $c= 2.20$.}
 \renewcommand{\arraystretch}{0.75}
 \renewcommand{\topfraction}{0.6}
  \label{DLossAndDMin}
\end{center} 
\end{figure}

Our goal of reducing the noises in LIGO is constrained by the need to keep the diffraction loss at nearly 1 ppm.  The hyperboloidal beams have two parameters, $D$ and $\alpha$.  $D$ is roughly the width of the beam, and $\alpha$ is the shape.  The duality relation reduces the range of $\alpha$ that we need to consider.  We focus our attention on $\pi/2 \leq \alpha \leq \pi$ since the intensity profiles are identical to those in the range from $0$ to $\pi/2$ but the mirrors are nearly concentric as needed to decrease the tilt instability.  The Mesa profile obtained for $\alpha = \pi$ has the flat-top shape required to decrease the thermal noises.  As $\alpha$ goes toward $\pi / 2$ the beam becomes more rounded, losing the flat top and sharp decay near the edge.  We have shown that $D \to \infty$ the beam becomes a Gaussian whose width is $w_0 \sqrt{1 + \cos^2 \alpha} / \sin \alpha$.  For $\alpha$ near $\pi$, this Gaussian is nearly infinitely wide, and at $\alpha = \pi / 2$ the beam becomes the minimal Gaussian of width $w_0$.  From the clipping approximation we can estimate that at $\alpha \approx 0.247 \pi$ and $\alpha \approx 0.753 \pi$, $D = \infty$ the diffraction loss is about 1 ppm.  We do not have to consider any values of $\alpha$ between these since the widest beam consistent with the diffraction loss constraint would be the $D = \infty$ Gaussian.

\begin{table}[h]
\begin{center}
\begin{tabular}{|>{\footnotesize}p{1.0in}>{\footnotesize}p{1.0in}|}
\hline
$\alpha = \pi$ &$ D =  9.62$ cm \\ \hline \hline
Diffraction Loss & Phase   \\ \hline
1.0 ppm & 0.0  \\ 
121.7 ppm & -1.5104 \\ 
182.3 ppm & -0.5835  \\
334.8 ppm &-2.6677 \\ 
7941.6 ppm &2.2904  \\
45401.8 ppm &0.8325  \\  \hline \hline
$\alpha =  0.95 \pi$ &$ D = 9.71 $ cm \\ \hline \hline
Diffraction Loss & Phase   \\ \hline
1.0 ppm & 0.0 \\
136.2 ppm & -1.5079 \\
195.3 ppm & -0.5802 \\
302.3 ppm &-2.6630 \\
7225.3 ppm & 2.2976 \\
44104.7 ppm & 0.8422 \\ \hline \hline
$\alpha =  0.90 \pi$ &$ D =  11.01 $ cm \\ \hline \hline
Diffraction Loss & Phase   \\ \hline
1.0 ppm & 0.0 \\
18.6 ppm & -0.4797 \\
951.8 ppm & 1.3048 \\
3400.8 ppm &  2.7653 \\
5870.7 ppm &  -2.3322 \\
61706.3 ppm & 1.4493 \\ \hline
\end{tabular}
\end{center}
\caption{The diffraction loss and phase separation for eigenvalues with losses less than 10 percent, in three different hyperboloidal configurations with 1 ppm loss in the fundamental mode. Only axisymmetric modes are shown.}
\label{tablephases}
\end{table}

\subsection{Finite Mirror Effects}
If the mirrors were infinite in extent, no light would propagate off of the mirror and there would be no diffraction loss, giving eigenvalues of unit magnitude.  The clipping approximation assumes that the beam is the one supported by infinite mirrors.  This is not the case as diffraction also causes the beam profile to change.  The propagator is a more accurate calculation because it finds the precise beam profile supported by the mirrors.  The clipping approximation is typically an underestimate of the diffraction loss \cite{O'Shaughnessy:2004qh,D'Ambrosio:2004wz}.  In Fig.\ \ref{DLossvsR}a, we show that this is indeed usually the case.  However, for some ranges of beam width $D$ there is an anomalously low diffraction loss below the clipping approximation.

To study this effect, we varied the mirror radius and computed the diffraction loss as a function of $D$.  The mirrors that we study in this paper have radius $R = 17$ cm, so we compared with $R = 16$ cm and $R = 18$ cm.  The local minimum becomes narrower and shallower for increasing mirror radii.  This is suggestive of a finite mirror effect that will disappear when the mirror radius is significantly larger than the radius of the beam.

Fig.\ \ref{DLossvsR}b shows the deviation of the beam from the infinite mirror Mesa beam.  A concentric Mesa ($\alpha = \pi$) beam with $D = 10.67$ cm is chosen.  This is the location of a local minimum of the diffraction loss with respect to $D$.  The beam intensity is computed directly from the eigenvector of the propagator.  The intensity is normalized by integrating over the mirror rather than over the entire mirror plane as with the infinite mirror case.  However, this only causes a fractional error in the normalization on the order of $10^{-6}$.  As shown in the figure, the finite mirror causes deviations from the infinite mirror beam; the plateau of the beam is less flat than expected.  When the radius of the mirror is increased, the deviations retain their shape but decrease in size.  For $R = 20$ cm, the beam is very close to the infinite mirror expectation.  There is still a difference near the outer edge of the beam, causing the intensity to decay more quickly with radius than in the infinite mirror case.

Our numerical results suggest that the anomalous diffraction loss is related to the deviation of the beam from the ideal Mesa shape.  As $D$ increases, the clipping approximation predicts a smooth increase in diffraction loss due to the widening beam.  Finite mirror effects increase with the ratio $D / R$, so they alone do not explain this unexpected behavior.  For the values of $D$ that yield an anomalous diffraction loss, the variations around the plateau (Fig.\ \ref{DLossvsR}b) have an organized shape with an approximate wavelength of $w_0$.  The variations in these cases have a shape such that they alter the falloff of the beam at the edge, i.e. in the last two centimeters of the mirror.  This has an obvious beneficial effect on the diffraction loss.

\begin{figure}[h]
\begin{center}
\leavevmode
\epsfxsize=235pt
\epsfbox{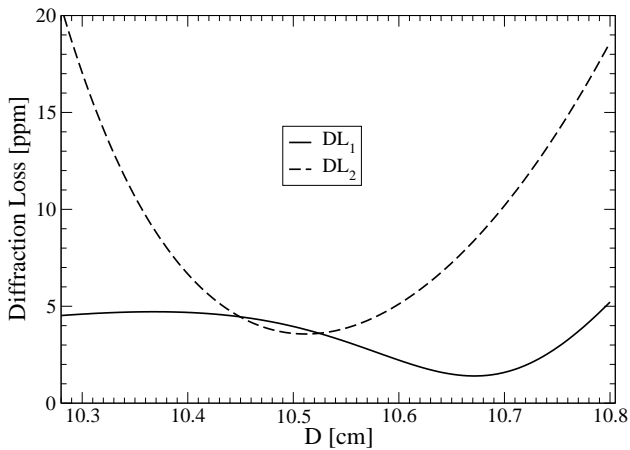}
\caption{The diffraction losses of the fundamental mode and the first excited mode are plotted as a function of $D$ for the $\alpha = \pi$ Mesa configuration.  The two curves cross due to finite mirror effects causing an anomalous diffraction loss for the first excited mode.}
 \renewcommand{\arraystretch}{0.75}
 \renewcommand{\topfraction}{0.6}
  \label{Crossing}
\end{center} 
\end{figure}

We expect that the fundamental mode of the cavity (the hyperboloidal shape, with no nodes), should have the lowest diffraction loss.  However, this is not the case for all choices of $\alpha$ and $D$.  Surprisingly, the diffraction loss of the first excited axisymmetric mode can decrease below that of the fundamental mode.  This occurs, for $\alpha = \pi$, for a small range of $D$ around 10.5 cm, with diffraction losses around 4 ppm.  The arguments of two eigenvalues remain well separated.  Fig.\ \ref{Crossing} shows the crossing of the two diffraction loss curves.

To investigate the cause of this crossing, we choose the specific value $D = 10.52$ cm and increase the mirror radius to 18 cm.   The diffraction loss of the fundamental mode decreases from 3.7 ppm to 0.4 ppm, while the first excited mode only decreases from 3.6 ppm to 2.3 ppm so the diffraction loss of the fundamental mode is now lowest.  Further increasing the mirror radius to 20 cm causes the losses of the second and third excited modes to cross.  Despite changes of the mirror radius, the arguments of the eigenvalues change by less than a percent (we are only considering eigenvalues with losses less than 10,000 ppm because otherwise they would dissipate too quickly to be of interest).  This dependence on mirror radius tends to confirm that this is a finite mirror effect.  Having demonstrated that this effect can have substantial and beneficial effects on the diffraction loss, we now turn our attention to studying the parameter space of hyperboloidal beams in more detail.  Further studies of the precise cause of the anomalous losses may want to focus on the deviation of the beam from its theoretical expectation, as well as the behavior with changing mirror radius.

\subsection{Noises for fixed D}
\begin{figure}[h]
\begin{center}
\leavevmode
\epsfxsize=235pt
\epsfbox{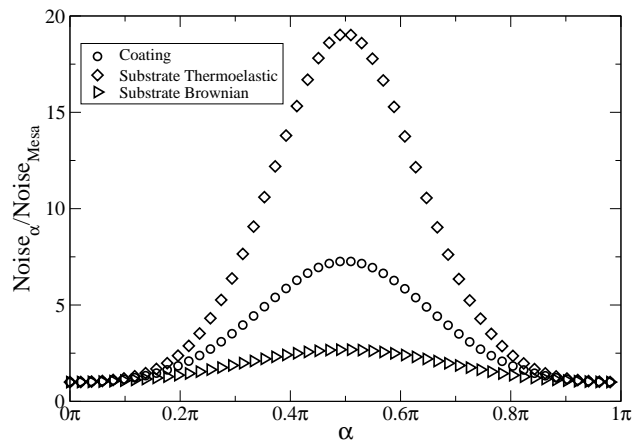}
\caption{The noise ratios $\rm{Noise}_\alpha/\rm{Noise}_{\rm{Mesa}} - 1$ for three types of noise are shown as a function of $\alpha$ for fixed $D = 10$ cm.  The minimal Gaussian $\alpha = \pi/2$ has the highest noise and Mesa ($\alpha =0, \pi$) has the lowest noise.}
 \renewcommand{\arraystretch}{0.75}
 \renewcommand{\topfraction}{0.6}
  \label{NoiseFixD}
\end{center} 
\end{figure}

The width of the beam increases with increasing $D$, which averages the fluctuations over more of the mirror surface and therefore decreases the noise.  It is less clear how the noise will behave when $\alpha$ is changed.  We begin by fixing $D = 10$ cm and ignoring the diffraction loss constraint. Fig.\ \ref{NoiseFixD} shows that all three types of noise increase as $\alpha$ moves away from $0$ and $\pi$.  Substrate thermoelastic noise is most affected by changing $\alpha$, followed by the coating noises (recall that both types of coating noise follow the same scaling law).

As $\alpha$ is decreased from $\pi$, the noises increase if $D$ is kept fixed.  At the same time, the diffraction loss decreases.  If we keep the diffraction loss fixed, the $D$ can be increased as $\alpha$ decreases toward $\pi / 2$.  Widening the beam tends to decrease the noise, which partially offsets the increase from changing $\alpha$.  We fixed the loss at 1.4 ppm and found that the noise still increases for beams other than Mesa.  We expect that for larger diffraction losses this result will still hold.  However, at 1 ppm the anomalous behavior of the diffraction loss due to finite mirror effects changes this conclusion.

\subsection{Hyperboloidal beams with 1 ppm Diffraction Loss}
As $\alpha$ decreases from $\pi$ toward $\pi / 2$, the beam loses its flat top and sharp falloff, and approaches the minimal Gaussian.  Also, the $D$ corresponding to the local minimum in diffraction loss increases, and the local minimum becomes deeper and wider.  Fig.\ \ref{DLossvsD} shows the diffraction loss versus $D$ for three values of $\alpha$, while Fig.\ \ref{DLossAndDMin} gives $D$ and diffraction loss at the local minimum for a range of $\alpha$.  The local minimum of the diffraction loss for a Mesa beam ($\alpha = \pi$) is at $D = 10.67$ cm, and has $1.4$ ppm.  As discussed above, the Mesa is has the lowest noise of the hyperboloidal family for this diffraction loss.

\begin{figure}
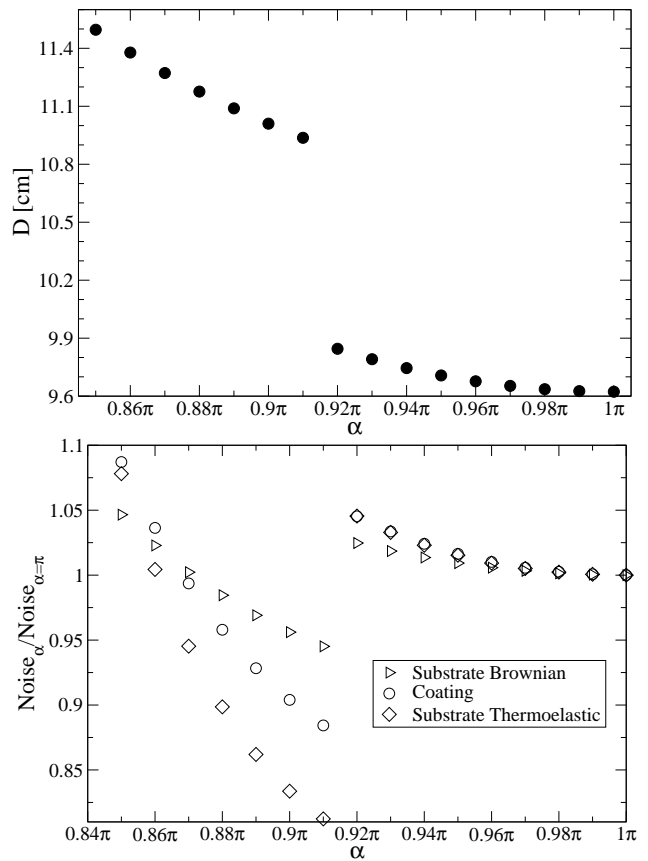

\begin{center}
\leavevmode
\epsfxsize=235pt
\epsfbox{fig7a.eps}
\newline
\epsfxsize=235pt
\epsfbox{fig7b.eps}
\caption{ (a) The largest values of $D$ giving a strict 1 ppm diffraction loss are shown as a function of $\alpha$. The discontinuity is caused by finite mirror effects as explained in the text. (b) The corresponding values of the noise, normalized so that the noises for the 1 ppm Mesa ($\alpha = \pi$, $D = 9.62$ cm) are all equal to 1.  The discontinuity in allowed maximum $D$ leads to a sharp drop in noise at $\alpha = 0.91 \pi$.}
 \renewcommand{\arraystretch}{0.75}
 \renewcommand{\topfraction}{0.6}
  \label{OnePPMNoises}
\end{center} 
\end{figure}

If a diffraction loss of strictly $1$ ppm is required, the beam must be reduced to a width of $D = 9.62$ cm in the Mesa configuration.  Alternatively, we can consider other values of $\alpha$.  The local minima are displayed in Fig.\ \ref{DLossAndDMin}, which gives the values of $D$ as well as the corresponding diffraction losses.  Note that for $\alpha = 0.9 \pi$, the diffraction loss at the local minimum is now below 1 ppm.  Fig.\ \ref{OnePPMNoises}a shows the maximum $D$ that yields a 1 ppm loss, as $\alpha$ is varied.  There is a discontinuity between $\alpha = 0.91 \pi$ and $0.92 \pi$ because below $0.92 \pi$ the diffraction loss at the local minimum is below 1 ppm.  The noises therefore drop substantially when $\alpha = 0.91 \pi$ as in Fig.\ \ref{OnePPMNoises}b.  The coating noise decreases by 12\% and the substrate thermoelastic by 19\%, relative to the 1 ppm Mesa beam.  A strict requirement of 1 ppm diffraction loss therefore combines with the finite mirror effects to make the $\alpha = 0.91 \pi$, $D = 10.94$ cm configuration the best choice.

\subsection{Correcting for Finite Mirror Effects}

Restructuring the mirror to specifically account for finite mirror effects allows us to increase $D$ in hyperboloidal beams (thereby reducing thermal noise), while keeping within given diffraction loss constraints. The restructured beams can reduce the diffraction loss by a factor of 30 to 100, allowing a wider beam. For the Mesa case this allows for a net noise reduction of ~30\% for the beam satisfying the 1ppm diffraction loss constraint.

As noted above, the original Mesa beam used to construct the mirror is infinite in extent. The mirror is designed to be a phasefront of the theoretical beam. Since the mirror is actually finite, for $D \sim R$ there can be substantial effects due to missing light that was incident on the mirror plane outside the mirror radius. To account for these finite mirror effects we reconstruct the mirror, with the goal of  making the phase of the first eigenbeam constant at the mirror surface, rather than the phase of the idealized infinite beam.

The propagator formulation allows us to explicitly calculate the phase of the eigenbeams, as a function of $r$. As the mirror deviation from the fiducial spheroid $h(r)$ enters into the calculation of phase through the propagator, we use an iteration scheme to adjust the mirror to match the eigenbeam phasefront motivated by the argument of the propagator: 
\bea
\arg[\mathcal{K}(r,r')] &\simeq& \pi/2 + k_0 \left(h(r) + h(r') - L\right)  \label{argprop}\\
&\qquad& - k_0 \cos \alpha \left( \frac{r^2}{2L} + \frac{r'^2}{2L}\right)~.\nonumber
\eea
We see in Eq. \eqref{argprop} that if the phase of the eigenbeam is too large at some radius $r$, reducing the value of $h(r)$ should act to roughly reduce the phase of the new eigenbeam.

With this motivation, we apply the simple iteration scheme:
\beq
h_{i+1}(r) = h_{i}(r) - c \times \arg\left[V^{(1)}_{i}(r)/V^{(1)}_{i}(0)\right] ~,
\eeq
where the $i$ denotes the $i^{\rm th}$ iteration, $V^{(1)}_{i}(r)$ is the first eigenbeam for the mirror with deviation $h_i(r)$, and $c > 0$ is an arbitrary constant less than unity, chosen to prevent overshoot. 

\begin{figure}
\begin{center}
\leavevmode
\epsfxsize=235pt
\epsfbox{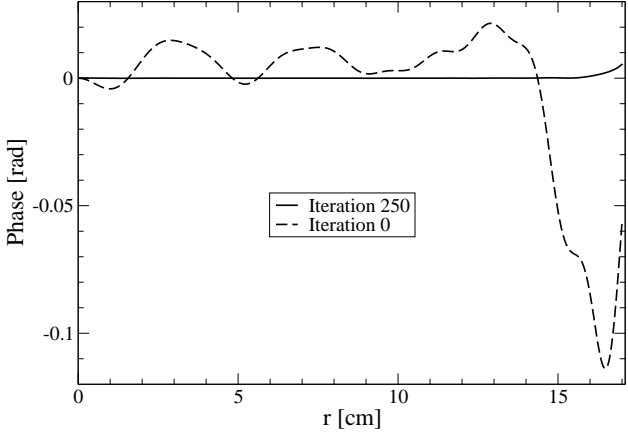}
\caption{The phase of the fundamental eigenbeam as a function of radius is shown for iteration zero and 250. It can be seen that the iteration scheme drives it closer to zero as expected.}
 \renewcommand{\arraystretch}{0.75}
 \renewcommand{\topfraction}{0.6}
  \label{Phase1stEig}
\end{center} 
\end{figure}

\begin{figure}[h]
\begin{center}
\leavevmode
\epsfxsize=235pt
\epsfbox{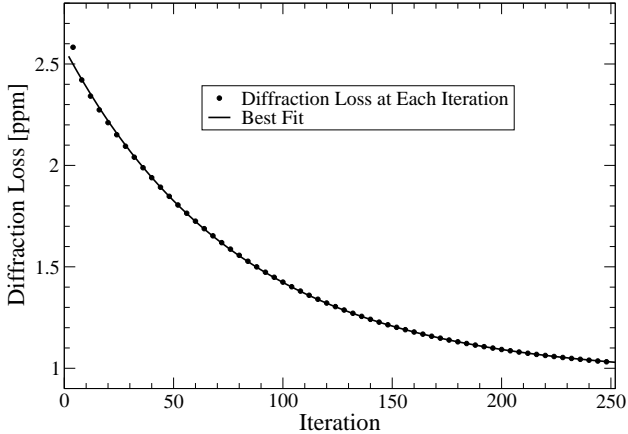}
\caption{The diffraction loss is shown as function of iteration number for a $\alpha = \pi$ Mesa configuration with $D=11.35$ cm. The original beam (not shown) has diffraction loss at 46.5 ppm, and it can be seen that after a few iterations the diffraction loss begins to converge to an exponential with lower bound $\sim 1$ ppm. The best fit exponential is given by $0.96 + 1.616 \exp(- 0.013 i)$ ppm, where $i$ is the iteration number.}
 \renewcommand{\arraystretch}{0.75}
 \renewcommand{\topfraction}{0.6}
  \label{Iterations}
\end{center} 
\end{figure}

\begin{table}[h]
\begin{center}
\begin{tabular}{|>{\footnotesize}p{1.0in}>{\footnotesize}p{1.0in}|}
\hline
$\alpha = \pi$ &$ D =  11.35$ cm (Uniterated) \\ \hline \hline
Diffraction Loss & Phase   \\ \hline
46.5 ppm & 0.0 \\ 
128.5 ppm & -0.4313 \\
341.7 ppm &-1.1895 \\ 
10530.8 ppm& -2.1470 \\
38445.0 ppm  & 3.0277 \\ \hline \hline
$\alpha = \pi$ &$ D =  11.35$ cm (Iterated) \\ \hline \hline
Diffraction Loss & Phase   \\ \hline
1.0 ppm  & 0.0 \\
320.6 ppm & -0.4319 \\
1100.1 ppm &-1.1920 \\
26167.4 ppm & -2.1593 \\
66808.7ppm & 2.9873 \\ \hline \hline
$\alpha = 0.9 \pi$ &$ D =  11.87$ cm (Uniterated) \\ \hline \hline
Diffraction Loss & Phase   \\ \hline
43.5 ppm & 0.0 \\ 
205.0 ppm & -0.4173 \\
371.2 ppm &-1.1665 \\ 
10626.6 ppm& -2.1082 \\
50723.6 ppm  & 3.0878 \\ \hline \hline
$\alpha = 0.9 \pi$ &$ D =  11.87$ cm (Iterated) \\ \hline \hline
Diffraction Loss & Phase   \\ \hline
1.0 ppm  & 0.0 \\
366.6 ppm & -0.4185 \\
2013.4 ppm &-1.1701 \\
32628.3 ppm & -2.1267 \\
87359.8 ppm & 3.0335 \\ \hline \hline
\end{tabular}
\end{center}
\caption{The phase separation for the axisymmetric ($m=0$) modes with diffraction losses less than 10\%, both before and after the iteration scheme is applied. The phases do not change significantly as the mirror is iterated. The absolute value of the eigenvalues with nonzero phase (and hence the diffraction loss) increase upon iterating. Preliminary calculations show that non-axisymmetric modes have diffraction losses increased by the iteration process while the phases change by no more than 5\%.}
\label{tablephasesDave}
\end{table}

\begin{table}[h]
\begin{center}
\begin{tabular}{|>{\footnotesize}p{0.3in}>{\footnotesize}p{0.4in}>{\footnotesize}p{0.6in}>{\footnotesize}p{0.6in}>{\footnotesize}p{0.8in}|}
\hline
$\alpha$ &$D$ [cm] & Coating Noise & Substrate Brownian Noise & Substrate Thermoelastic Noise\\ \hline \hline
$\pi$ &$ 11.35$ & 0.72 & 0.84 & 0.63 \\ 
$0.9\pi$ &$11.87$ & 0.80 & 0.90 & 0.69 \\  \hline
\end{tabular}
\end{center}
\caption{The coating, substrate Brownian and substrate thermoelastic noise are 
displayed after the iteration process. The diffraction loss is kept constant
at 1 ppm. The noises are normalized to noises of the original ($\alpha = \pi$) Mesa (with $D = 9.62$ cm)
that gives the 1 ppm diffraction loss.  It can be seen that the iteration scheme lowers the
noise by about 30\% for $\alpha = \pi$ by allowing larger $D$ for the same 1 ppm diffraction loss.}
\label{tablenoise}
\end{table}

\begin{figure}
\begin{center}
\leavevmode
\epsfxsize=235pt
\epsfbox{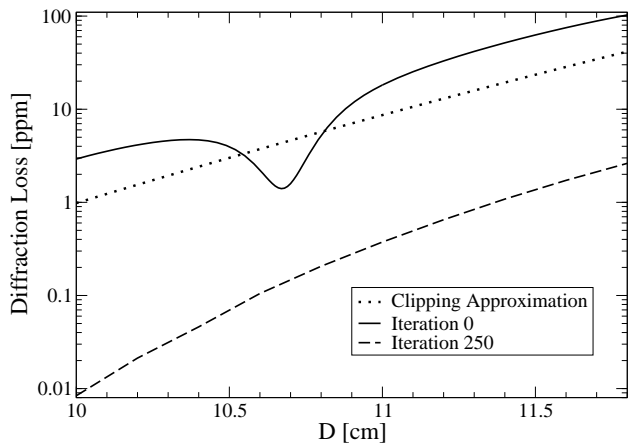}
\caption{The diffraction loss in ppm calculated using the clipping approximation is compared to that using the propagator eigenvalues for the iterated and original mirrors as a function of $D$. As before, the configuration studied is $\alpha = \pi$ Mesa with $R = 17$ cm.  The iteration process lowers the diffraction loss by a factor of 30 to 100.}
 \renewcommand{\arraystretch}{0.75}
 \renewcommand{\topfraction}{0.6}
  \label{DaveDiffList}
\end{center}
\end{figure}

This iteration scheme successfully reduces the relative phase of the eigenbeam, as shown in Fig \ref{Phase1stEig}. Optimizing the mirror surface to match the phasefront of the primary eigenbeam also acts to reduce the diffraction loss for that mode in general, with the iteration scheme providing convergence towards an apparent lower bound for the diffraction loss, while increasing the diffraction loss for other higher-order eigenbeams. This lower bound increases with $D$ (Fig \ref{DaveDiffList}).

This diffraction loss is plotted against the iteration number for the Mesa ($\alpha = \pi$) case with $D = 11.35$ cm in Fig \ref{Iterations}. The iteration scheme is shown to lower the diffraction loss for this $D$ from 46.5 ppm to a $\sim 1$ ppm lower bound, satisfying the required design constraint. The diffraction losses of higher order modes are more than doubled in the iterated case as illustrated in Table \ref{tablephasesDave}.

\begin{figure}
\begin{center}
\leavevmode
\epsfxsize=235pt
\epsfbox{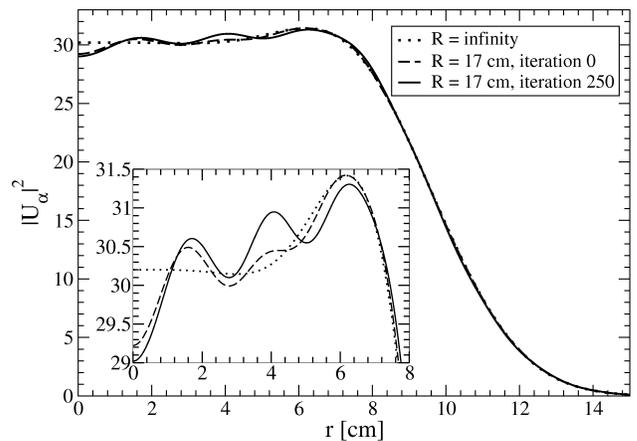}
\caption{The intensity profile $|U_\pi|^2$ for the mirror with $R=\infty$, $R = 17$ cm uniterated and $R = 17$ cm at iteration 250 are compared for the $\alpha = \pi$ Mesa configuration with $D = 11.35$ cm. The finite mirror effects induce oscillations in the intensity profile that do not disappear when the mirror is corrected.  An inset shows the central 8 cm `plateau' of the beam in detail.}
 \renewcommand{\arraystretch}{0.75}
 \renewcommand{\topfraction}{0.6}
  \label{DaveIntensity}
\end{center} 
\end{figure}

The beam for the iterated mirror with $D = 11.35$ cm is close to the original Mesa, but with variations in the central plateau of relative amplitude $\sim 1/30$ and variations of radial wavelength $\sim w_0$ (Fig \ref{DaveIntensity}). This seems to be an unavoidable consequence of a finite $R$, as even the original eigenbeam has roughly similar features. Despite changing the variations in the plateau of the beam intensity the process of iteration does not significantly affect the noises computed using Eq. \eqref{GLNoise1}.

\begin{figure}
\begin{center}
\leavevmode
\epsfxsize=235pt
\epsfbox{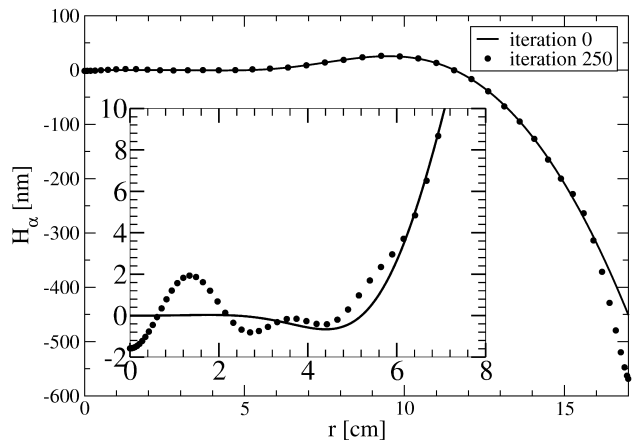}
\caption{The correction to the mirror $H_\pi$ at iteration 0 and at iteration 250 are compared for the $\alpha = \pi$ Mesa configuration with $D = 11.35$ cm, $R = 17$ cm. The iteration scheme introduces some bumps on the mirror of the size $\sim 2$ nm.  The inset shows the central 8 cm of the mirror in more detail.}
 \renewcommand{\arraystretch}{0.75}
 \renewcommand{\topfraction}{0.6}
  \label{DaveMirror}
\end{center} 
\end{figure}

Similarly the iterated mirror has variations of the central mirror shape of similar radial scale, with amplitude on the order of 2 nm, shown in Fig \ref{DaveMirror}. The most significant feature of the iterated mirror is the inward tilting of the outer edge of the mirror, though preliminary studies show that the primary contribution to reducing the diffraction loss is due to the variations near the center of the mirror.

Reformulating the mirror to account for finite mirror effects allows us to increase the $D$ parameter of the beam from $9.62$ cm to $11.35$ cm for a concentric Mesa beam while still maintaining a 1ppm diffraction loss. Though this design may introduce more complications in the construction of the mirror itself, it allows a significant reduction in noise by broadening the beam. This iteration scheme can also be used for other values of $\alpha$, as shown in Table III, where the iterated mirror for $\alpha = 0.9 \pi$ has a diffraction loss lower bound of 1 ppm for $D  = 11.87$ cm. However, we find that $\alpha = \pi$ is optimal for noise reduction. (See Table \ref{tablenoise}.)

\section{Conclusions}
In this paper, we studied thermal noise and diffraction loss for the hyperboloidal family of light beams and mirror shapes in detail for the first time. This family had been initially proposed to unify the concentric ($ \alpha = \pi$) and nearly-flat ($\alpha = 0$) Mesa configurations through variations of the twist angle $\alpha$. In this paper we also presented an analytic proof that Gaussian beams are a limiting case of the hyperboloidal beam as $D \to \infty$. This was previously conjectured in Ref. \cite{BT}. We developed a pseudo-spectral code both fast and accurate enough to calculate the diffraction loss directly from the beam propagator. We find that the finite radius of the mirror causes beam shapes to deviate significantly from the infinite-mirror theoretical expectations. This causes a previously unnoticed local minima in the diffraction loss that can be exploited to find a natural beam width $D$ for the current diffraction loss constraints of about 1 ppm. For an $\alpha = \pi$ Mesa beam a local minimum occurs at $D = 10.67$ cm giving 1.4 ppm diffraction loss for a mirror of radius $R = 17$ cm. If one requires a strict enforcement of the 1 ppm diffraction loss we show that a hyperboloidal beam with $\alpha = 0.91 \pi$ and $D = 10.94$ cm has lower noise than that of the $\alpha = \pi$ Mesa with 1 ppm diffraction loss. The coating noise changes by about 12\% and the substrate thermoelastic noise and substrate Brownian noise change by 5\% and 19\%, respectively.

We also propose new mirror and beam shape configurations that explicitly account for finite mirror effects by reformulating the mirror surface to coincide with the phasefront of the primary eigenbeam. These beams reduce the diffraction loss by more than an order of magnitude for the range of $D$ considered here (between $10$ cm and $11.8$ cm). This allows the use of wider beams for the same diffraction loss constraints on the primary eigenmode, while the diffraction losses of higher order modes (both axisymmetric and non-axisymmetric) are increased. We are able to widen the $\alpha = \pi$ Mesa beam for a $R = 17$ cm mirror while keeping the diffraction loss fixed at 1 ppm from a width of $D = 9.62$ cm to $D = 11.35$ cm. This lowers the coating thermal noise by about 30\% (compared to the smaller $D$ Mesa) and the other noises (substrate Brownian and substrate thermoelastic noise) by comparable factors. However, feasibility of the construction of the mirrors must also be taken into account.

The non-iterated beams taking advantage of the local minimum in diffraction loss discussed above are supported by strict hyperboloidal or Mesa mirrors, which would be no harder to make than the current Mesa designs,  and would still lower the coating thermal noise by 12\%.  If one is to consider the more ambitious goal of lowering the coating thermal noise by 28\% while using a beam that is very similar to Mesa through the iteration scheme described, the limitations on mirror manufacturing errors are likely to be more stringent, but still less than the currently considered conical beams \cite{MihaiYanbei,MihaiThesis}. In addition, the methods developed here for reducing the diffraction loss of the Mesa beam may be applied in the case of the conical beams previously considered \cite{MihaiThesis, MihaiYanbei}. The phase fronts of conical beams considered there have not been optimized to match the finite mirror surface. 

Recently, parametric instability \cite{PI1,PI2,PI3,PI4} was found to be a serious problem in Advanced LIGO. Choosing $D$ at the minimum of the diffraction loss curve of the hyperboloidal beams increases the diffraction loss of the higher eigenmodes, thus in principle somewhat improving the parametric instability. The effect is most pronounced in the case of the iterated mirrors which also have the most drastic thermal noise reduction.

\section*{Acknowledgements}
We thank Yanbei Chen for guidance, useful discussion and for foreseeing many of the issues presented in this paper. We are grateful to Geoffrey Lovelace for useful discussions regarding the noise calculations.  Our calculations make use of the GNU Scientific Library \cite{GSL}, IT++ \cite{ITPP}, and the AMD Core Math Library \cite{ACML}. We thank our advisors James York, \'Eanna Flanagan, Ira Wasserman, Saul Teukolsky, Dong Lai and Barry Barish for support and encouragement.
This work was partially supported by funds from the following sources: Sofja Kovalevskaja Programme from the Alexander Von Humboldt Foundation, NSF grants AST-0707628, AST-0606710 and PHY-0652952. APL and RB were supported to attend the GR18 / Amaldi7 conference, which was helpful to this work, by a combination of Cornell, NSF and IUPAP/ISGRG travel grants.

\end{document}